\documentclass[notitlepage, floatfix, tightenlines, showpacs, showkeys, nofootinbib, aps]{revtex4-1}

\usepackage{bm}
\usepackage{amsmath}
\usepackage{mathptmx}
\usepackage{amsfonts}
\usepackage{amssymb}
\usepackage[pdftex]{graphicx}
\usepackage{color}
\usepackage{balance}
\usepackage{mathptmx}
\RequirePackage[colorlinks,citecolor=darkred,urlcolor=deepmagenta,linkcolor=darkblue,bookmarks=false]{hyperref}
\usepackage{subfig}
\usepackage[toc,title]{appendix}
\usepackage[misc]{ifsym}
\usepackage{adjustbox}
\usepackage{multirow}
\usepackage{makecell}

\allowdisplaybreaks

\definecolor{darkblue}{rgb}{0.0, 0.0, 0.62}
\definecolor{deepmagenta}{rgb}{0.7, 0.01, 0.7}
\definecolor{darkred}{rgb}{0.55, 0.0, 0.0}
\begin{document}

\title{Naked singularity in $4D$ Einstein-Gauss-Bonnet novel gravity: Echoes and (in)-stability}
\author{Avijit Chowdhury}
\email[\Letter \hspace{0.1cm}]{avijit.phy@iitb.ac.in}
\affiliation{Department of Physics, Indian Institute of Technology Bombay, Mumbai 400076, India}%
\author{Saraswati Devi}
\email[\Letter \hspace{0.1cm}]{sdevi@iitg.ac.in}
\affiliation{Department of Physics, Indian Institute of Technology Guwahati, Guwahati,
Assam 781039, India}%
\author{Sayan Chakrabarti}
\email[\Letter \hspace{0.1cm}]{sayan.chakrabarti@iitg.ac.in}
\affiliation{Department of Physics, Indian Institute of Technology Guwahati, Guwahati,
Assam 781039, India}%

\begin{abstract}
We study the stability of an asymptotically flat, static, spherically symmetric naked singularity spacetime in the novel four dimensional Einstein-Gauss-Bonnet (EGB) gravity. The four dimensional EGB black hole for large enough values of the coupling parameter leads to such a naked singularity. The stability and the response of the spacetime is studied against the perturbations by test scalar, electromagnetic and Dirac fields and the time evolution of these perturbations were observed numerically. Implementing a null Dirichlet boundary condition near the singularity, we observed that for $l=1$ modes of scalar, electromagnetic perturbation and $l=0,1$ modes of Dirac perturbation the time-domain profile give rise to distinct echoes. However, as the coupling constant is increased, the echoes align and the QNM structure of the 4D-EGB naked singularity-spacetime becomes prominent. For higher values of the multipole number, the spacetime becomes unstable, thereby restricting the parameter space for the coupling parameter.    
\end{abstract}
\balance
\maketitle
\sloppy
\section{Introduction} \label{sec:intro}
Even after a century of its inception, Einstein's theory of general relativity (GR) is still the most widely accepted macroscopic theory of gravity. Despite its enormous success in both the weak field and strong field regime~\cite{will2014LRR,ishak2019LRR}, there is still no consistent way to connect the macroscopic theory of GR to a quantum field theory. Apart from this, GR does not give any satisfactory answer to the problem of local energy momentum conservation. It predicts space-time singularity and such singularities do have mathematical problems of their own. It is believed that singularities can not represent any physical object in nature. This led to the quest for alternative theories of gravity that would reduce to GR as a low energy effective theory. 
Interestingly, Lovelock~\cite{lovelock1971JMP} proved that in four dimensions, GR is the only metric theory of gravity that gives symmetric, covariant second order field equations in terms of the metric tensor. Thus, one of the ways, to modify GR is to work in spacetimes with dimensionality other than four~\cite{clifton2012PR}. In this regard, perhaps, the most general class of theories are the Lovelock theories which give symmetric, covariant second order field equations in terms of the metric tensor in arbitrary spacetime dimensions (see Ref.~\cite{paddy2013PR} for an excellent review). The Lovelock lagrangian is given by,
\begin{equation}
\mathcal{L}=\sqrt{-g}\left(-2 \Lambda + R + \alpha \mathcal{G}+\cdots\right)~,
\end{equation}
where $\mathcal{G}\equiv R^2 - 4 R_{\mu \nu} R^{\mu \nu}+ R_{\alpha \beta \mu \nu}R^{\alpha \beta \mu \nu}$ is known as the Gauss-Bonnet combination and gives the leading order correction to the Einstein-Hilbert action with a cosmological constant $\Lambda$. The Gauss-Bonnet term gives nontrivial signatures in $D>4$ dimensions, but is a topological invariant in four dimensions~\cite{chern1945AM}. The Gauss-Bonnet term apart from being quadratic in curvature invariants is of wide theoretical interest both from the perspectives of string theory and gravity ~\cite{ferrara1996PRL, antoniadis1997NPB, zwiebach1985PLB, nepomechie1985PRD, callan1986NPB, candelas1985NPB, gross1987NPB}. Thus, one is intrigued by the idea of consistently incorporating the effect of the Gauss-Bonnet term in a four-dimensional theory of gravity that will lead to field equations different from GR, circumventing Lovelock's theorem. The first step in this direction was taken by Glavan and Lin~\cite{glavan2020PRL}, who rescaled the Gauss-Bonnet coupling constant, $\alpha \rightarrow \alpha/(D-4)$, which in the limit $D \rightarrow 4$ cancels the vanishing contribution of the Gauss-Bonnet term. Consequently several works appear in the literature which includes formulation of cosmological solutions \cite{Li:2020tlo, kobayashi2020JCAP}, spherical black hole solutions \cite{Kumar:2020uyz, Fernandes:2020rpa, Kumar:2020owy, Ghosh:2020syx}, solutions of star-like objects \cite{Doneva:2020ped}, radiating and collapsing solutions \cite{Ghosh:2020vpc,Malafarina:2020pvl}, extending to more higher-curvature
Lovelock theories \cite{Konoplya:2020qqh}, thermodynamic behaviour of black holes in such theories \cite{EslamPanah:2020hoj, HosseiniMansoori:2020yfj,Konoplya:2020cbv, Hegde:2020xlv} and the gravitational and physical properties of these objects  \cite{Guo:2020zmf, konoplya2020EPJC, Zhang:2020qew, Roy:2020dyy, NaveenaKumara:2020rmi,Liu:2020vkh, Heydari-Fard:2021ljh, Kumar:2020sag, Islam:2020xmy,Mishra:2020gce,devi2020EPJC, churilova2021AP}.
Despite all these, the regularization scheme used in this novel four dimensional Einstein-Gauss-Bonnet ($4D$-EGB) theory~\cite{glavan2020PRL} was found to be inconsistent on several grounds~\cite{gurses2020EPJC, gurses2020PRL, arrechea2021CPC, arrechea2020PRL, bonifacio2020PRD, ai2020CTP, mahapatra2020EPJC, hohmann2021EPJP, cao2021arxiv}, which led to the development of different versions of regularized (consistent) $4D$-EGB theories~\cite{lu2020PLB, kobayashi2020JCAP, fernandes2020PRD, hennigar2020JHEP, aoki2020PLB, fernandes2021PRD}. 

There are several reasons why the method proposed in \cite{glavan2020PRL} does not work. It was observed that the field equations of Einstein-Gauss-Bonnet theory defined in its most general form in  $D>4$ dimensions can be split into two different parts. One of the parts of these field equations always remains higher dimensional, making the limiting procedure of $D\to4$ non-trivial \cite{gurses2020EPJC, gurses2020PRL, arrechea2021CPC, arrechea2020PRL,mahapatra2020EPJC}. Tree-level graviton scattering amplitudes were studied in this regard, independently of the Lagrangian, and it was shown that the dimensional continuation and $D\to 4$ limiting procedure applied to Gauss-Bonnet amplitudes does not produce any purely new four dimensional Gauss-Bonnet gravitational amplitudes \cite{bonifacio2020PRD}. All these imply that the existence of $D\to 4$ limiting solutions does not mean the existence of a four dimensional theory as proposed in \cite{glavan2020PRL}. However, interestingly enough, the field equations of the different versions of the $4D$-EGB gravity~\cite{ lu2020PLB, kobayashi2020JCAP, fernandes2020PRD, hennigar2020JHEP, aoki2020PLB} admit the same static spherically symmetric black hole solution as was proposed first in \cite{glavan2020PRL}, and from here onwards we will refer it to as $4D$-EGB black hole. The stability and quasinormal modes of the asymptotically flat $4D$-EGB black hole against perturbation by scalar, electromagnetic, Dirac fields has been studied in~\cite{konoplya2020EPJC}. Following the $D\rightarrow 4$ regularization of the scalar and vector type gravitational perturbation of the higher dimensional Einstein-Gauss-Bonnet black hole~\cite{takahashi2010PTP1, takahashi2010PTP2}, Konoplya {\it et. al} showed that the asymptotically flat, de Sitter and anti-de Sitter black holes are unstable in the eikonal limit (large $l$) for large positive values of the Gauss-Bonnet coupling parameter~\cite{konoplya2020EPJC, konoplya2020PDU}. The quasinormal modes of the $4D$-EGB black hole in the asymptotically de Sitter and anti-de Sitter spacetime due to scalar, electromagnetic and Dirac perturbations has been studied in~\cite{devi2020EPJC, churilova2021AP}. The quasibound states of massless scalar, electromagnetic and Dirac fields in the asymptotically flat $4D$-EGB black hole and the associated stability problem has been studied recently in~\cite{vieira2021arxiv}.

The $4D$-EGB black hole for large enough values of the coupling constant $\left(\alpha>M^2\right)$ leads to a naked singularity, violating the Cosmic Censorship Conjecture~\cite{penrose1969RNC}. Gyulchev {\it et al.}~\cite{gyulchev2021EPJC} studied the image of thin accretion disk around the weakly naked (with a photon sphere) $4D$-EGB singularity and observed a series of distinctive bright rings in the central part of the image which are otherwise absent for $4D$-EGB black holes. However, for any astrophysical system to be observationally relevant it has to be sufficiently stable. This naturally begs for the answer to the question that whether such a spacetime with a central naked singularity is at all stable under perturbation? If so, how its response will be different from that of a standard $4D$-EGB black hole~\cite{konoplya2020EPJC, konoplya2020PDU, vieira2021arxiv}? 

In this work, we try to answer these questions by studying the response of the $4D$-EGB naked singularity-spacetime towards the perturbation by test fields. We probe the spacetime by test scalar, electromagnetic and Dirac fields. Such an analysis does not give preference to any particular version of the (consistent) $4D$-EGB theory and as such can be regarded as more general. We observe that for the $l=1$ mode of scalar and electromagnetic perturbations and for the $l=0,1$ modes of Dirac perturbation, when $\alpha\gtrsim M^2$ , the signature of the difference between the spacetime due to a black hole and the naked singularity is quite distinctly elucidated by the existence of echoes in case of the $4D$-EGB naked singularity space time. However, that is not the only interesting feature that we obtain. We also find that as the coupling constant is increased further, the echoes align and the QNM structure of the 4D-EGB naked singularity-spacetime ringdown becomes prominent. For higher values of the multipole number, the spacetime becomes unstable, thereby restricting the parameter space of $\alpha$. In this regard, it is worth mentioning that presence of echoes in the ringdown signal due to perturbing fields was already predicted in the case of Janis-Newman-Winicour naked spacetime, which has a surface like naked singularity at a finite radial distance~\cite{chowdhury2020PRD}. In general, echoes highlight the existence of horizonless compact objects and have been predicted to be present in the ringdown signals of worm holes, fuzzballs and other exotic compact objects~\cite{cardoso2017NA, saraswat2019JHEP, maggio2020PRD, cardoso_PRD_2016, mark_PRD_2017, konoplya_PRD_2019, chirenti_PRD_2020, churilova_CQG_2020, bronnikov_PRD_2020, roy_arxiv_2019}. The presence of echoes has also been associated with modified theories of gravity~\cite{dong2021PRD} and existence of higher dimensions~\cite{dey2020PRD, dey2021PRD}. Echoes in gravitational waves is also expected to bear signatures of quantum gravity via quantization of the black hole area~\cite{agullo2021PRL, chakravarti2021arxiv}, although, Coates \textit{et al.}~\cite{Coates:2021dlg} have argued differently. For a detailed review on echoes in gravitational wave, we refer the reader to the excellent review by Cardoso and Pani~\cite{cardoso2019LRR}.

The paper is organised as follows. In Sec~\ref{sec:background}, we briefly describe the background spacetime, both in the black hole and the naked singularity regime. Section~\ref{sec:perturbation} discusses the perturbation of the 4D-EGB naked singularity spacetime by test fields. Section~\ref{sec:timedomain} is dedicated to the time domain analysis of the perturbation equations and evaluation of the associated quasinormal mode frequencies. Finally, in Sec.~\ref{sec:conclusion}, we conclude with a summary and discussion of the results.\\
Throughout the paper, we employ units in which $G = c = 1$. 
\section{Background Spacetime}\label{sec:background}
The asymptotically flat, static, spherically symmetric $4D$-EGB 
\begin{equation}\label{eq:metric}
ds^2= - f (r) dt^2 + \frac{1}{f(r)} dr^2 + r^2 \left(d \theta^2 + \sin^2{\theta} \right),
\end{equation}
where
\begin{equation}\label{eq:f(r)}
f(r)=1+\frac{r^2}{2\alpha}\left( 1-\sqrt{1+\frac{8 \alpha M}{r^3}} \right)
\end{equation}
with $\alpha$ being a positive constant and $M$ being the ADM mass. 
The spacetime~\eqref{eq:metric} also appears as a solution to semi-classical Einstein's equation with Weyl anomaly and in the context of Einstein gravity with quantum corrections~\cite{cai2010JHEP, cai2014PLB, cognola2013PRD}. The uniqueness of the black hole solution~\eqref{eq:metric} in the scalar-tensor formulation of the 4D-EGB theories has been discussed in~\cite{clifton2021PRD} along with another branch of solution that leads to a naked singularity. 

The nature of the solution~\eqref{eq:metric} depends on the values of the dimensionless constant parameter $\gamma=\alpha/M^2$. For $\gamma$ in the range $\left[0,1\right]$, the spacetime defined by Eq.\eqref{eq:metric} represents a black hole of mass $M$, characterised by an outer event horizon at $r_+$ and an inner horizon at $r_-$, hiding a central curvature singularity at $r=0$, where
\begin{equation}
r_\pm=M\left(1\pm\sqrt{1-\gamma}\right).
\end{equation}
For $\gamma=1$, the line element~\eqref{eq:metric} corresponds to an extremal black hole characterised by a single horizon at $r_+=r_-=M$. However, for $\gamma>1$, the horizons cloaking the singularity cease to exist and the singularity at $r=0$ becomes globally naked~\cite{gyulchev2021EPJC}. It was shown that the Kretschmann scalar diverges at the location of the singularity at $r = 0$, however it does so in a slower rate than the Schwarzschild one. For $\gamma$ in the range $(1,3\sqrt{3}/4)$, the spacetime is surrounded by a photon sphere of radius $r_{ph}$ and the singularity is classified as  being ``Weakly naked'', whereas for $\gamma>3\sqrt{3}/4$ no such photon rings are present and the singularity is classified as ``strongly naked"~\cite{virbhadra2002PRD}.

\section{Perturbation by test fields}\label{sec:perturbation}
 To analyse the stability and ringdown signatures of the spacetime~\eqref{eq:metric} with a central naked singularity, we study the perturbation of the spacetime against test scalar, electromagnetic and Dirac fields.
\subsection{Scalar field}\label{subsec:scalar}
The dynamics of a massless test scalar field $\Psi$ propagating in the background~\eqref{eq:metric} is governed by the Klein-Gordon equation,
\begin{equation}\label{eq:KG}
\frac{1	}{\sqrt{-g}}\partial_\mu\left(\sqrt{-g} g^{\mu \nu} \partial_\nu \Psi_{scalar}\right)=0~.
\end{equation}
The spherical symmetry of the background spacetime allows us to separate out the angular dependence of the scalar field $\Psi$ as,
\begin{equation}\label{eq:scalar_decomp}
\Psi(t,r,\theta,\phi)=\frac{1}{r} \psi_{scalar}(t,r) Y_{l m} \left(\theta ,\phi \right),
\end{equation}
where $Y_{l m} \left(\theta, \phi \right)$ are the spherical harmonics of degree $l$ and order $m$.
Thus, Eq.~\eqref{eq:KG} can be rewritten as
\begin{equation}\label{eq:scalar_trts}
\frac{\partial ^2 \psi_{scalar}}{\partial t^2} -\frac{\partial ^2 \psi_{scalar}}{\partial r_*^2} + V_{scalar}(r) \psi_{scalar}=0
\end{equation}
where,
\begin{equation}\label{eq:v_scalar}
V_{scalar}(r)=f(r)\left(  \frac{l(l+1)}{r^2} + \frac{1}{r} \frac{d f(r)}{d r}\right),
\end{equation}
is the effective potential for scalar field perturbation
The coordinate $r_*$ is defined analogous to the tortoise coordinate of black holes,
\begin{equation}
dr_*=\frac{dr}{f(r)}.
\end{equation}
Close to the singularity, the coordinate $r_*$ varies linearly with $r$, such that the singularity is by definition at $r_*=0$.
\footnote{For numerical computations, we shifted the origin of the tortoise coordinate from $r=0$ to $r=0+\epsilon$, $\epsilon << 1$. Thus, the exact position of the singularity is excluded from the domain of numerical study but the effect of the singularity in terms of the divergence of the effective potential drives the dynamics of the test fields.}
\subsection{Electromagnetic field}\label{subsec:em}
The motion of a test electromagnetic field in a curved background is given by the equation
\begin{equation}\label{eq:em}
\frac{1}{\sqrt{-g}}\partial_{\mu}\left(\sqrt{-g} F_{\gamma \sigma} g^{\gamma \mu} g^{\sigma \nu}\right) =0,
\end{equation}
where $F_{\gamma\sigma}=\partial_{\gamma} A_{\sigma}-\partial_{\sigma} A_\gamma$ and $A_\mu$ is the four vector potential. The spherical symmetry of the background spacetime allows us to decompose the angular part of $A_\mu$ in terms of the vector spherical harmonics,
\begin{equation}\label{eq:vectordecom}
\begin{split}
&A_\mu (t,r,\theta,\phi)\\
&=\sum_{l,m}\left( 
\begin{bmatrix}
0\\
0\\
\frac{a^{lm}(t,r)}{\sin \theta}\partial_\phi Y_{lm}\\
-a^{lm}(t,r)\sin \theta \partial_\theta Y_{lm}
\end{bmatrix} 
+ 
\begin{bmatrix}
f^{lm} (t,r) Y_{lm}\\
h^{lm} (t,r) Y_{lm}\\
k^{lm} (t,r) \partial_{\theta} Y_{lm}\\
k^{lm} (t,r) \partial_{\phi} Y_{lm}
\end{bmatrix}
\right) 
\end{split}
\end{equation}
where the first term inside the summation is of odd parity, $\left(-1\right)^{l+1}$, while the second term is of even parity $\left(-1\right)^{l}$. Plugging Eq.~\eqref{eq:vectordecom} back into Eq.~\eqref{eq:em} one can arrive at a Schr\"{o}dinger like equation,
\begin{equation}\label{eq:em_trts}
\frac{\partial ^2 \psi_{em}}{\partial t^2} -\frac{\partial ^2 \psi_{em}}{\partial r_*^2} + V_{em}(r) \psi_{em}=0,
\end{equation}
where $\psi_{em}= a^{lm}$ for odd parity and $\psi_{em}=\frac{r^2}{l(l+1)}\left(\partial_t h^{lm}-\partial_r f^{lm} \right)$ and 
\begin{equation}\label{eq:v_em}
V_{em}(r)=f(r)\frac{l(l+1)}{r^2}
\end{equation}
is the effective potential for electromagnetic perturbation.
 
\subsection{Dirac field}\label{subsec:dirac}
The dynamics of a massless fermionic field $\Upsilon$ in a curved background is determined by the Dirac equation,
\begin{equation}\label{eq:dirac}
\gamma^\mu\left(\partial_\mu -\Gamma_\mu\right)\psi_{dirac}=0,
\end{equation}
where $\gamma^\mu$ are the coordinate dependent Dirac four-matrices and $\Gamma_\mu$ are the spin connections in the tetrad formalism. Following \cite{brill1957RMP}, we separate out the angular dependence and rewrite the covariant equations of motion Eq.~\eqref{eq:dirac} as 
\begin{equation}\label{eq:dirac_trts}
\frac{\partial^2 \psi^{\pm}_{dirac}}{\partial t^2} -\frac{\partial^2 \psi^{\pm}_{dirac}}{\partial r_*^2} + V^{\pm}_{dirac}(r) \psi^{\pm}_{dirac}=0,
\end{equation}
where 
\begin{equation}\label{eq:v_dirac}
V^{\pm}_{dirac}(r)=\frac{l+1}{r} f(r)\left(\frac{l+1}{r}\mp \frac{\sqrt{f(r)}}{r}\pm \frac{d\sqrt{f(r)}}{dr}\right)
\end{equation}
are the effective potential corresponding to the two chiralities labelled as `$+$' and `$-$'. However, for the background spacetime of the form ~\eqref{eq:metric}, one can transform the potential for opposite chiralities into one another following a Darboux transformation and hence they are isospectral. Thus, we will only consider the potential $V^+_{dirac}$~.

\begin{figure*}[]
\includegraphics[width=0.33\textwidth]{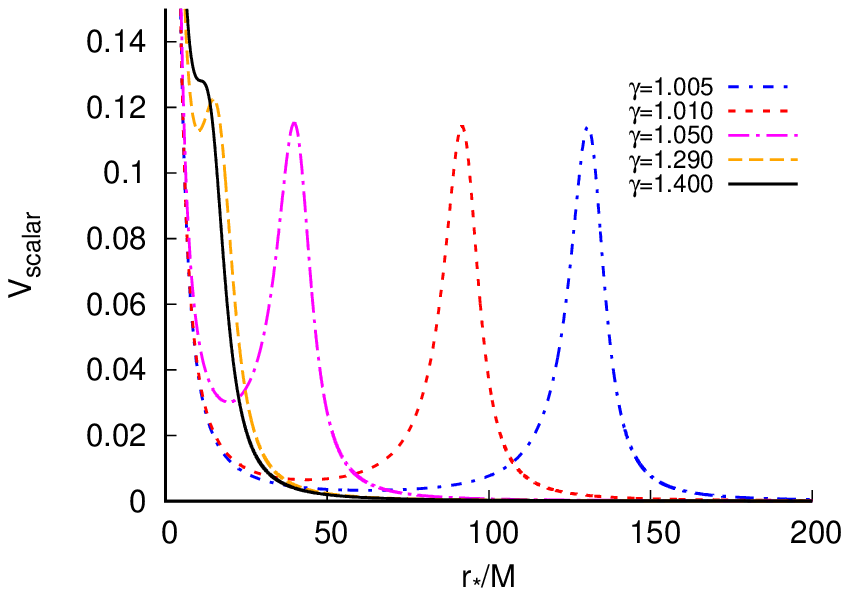}
\includegraphics[width=0.33\textwidth]{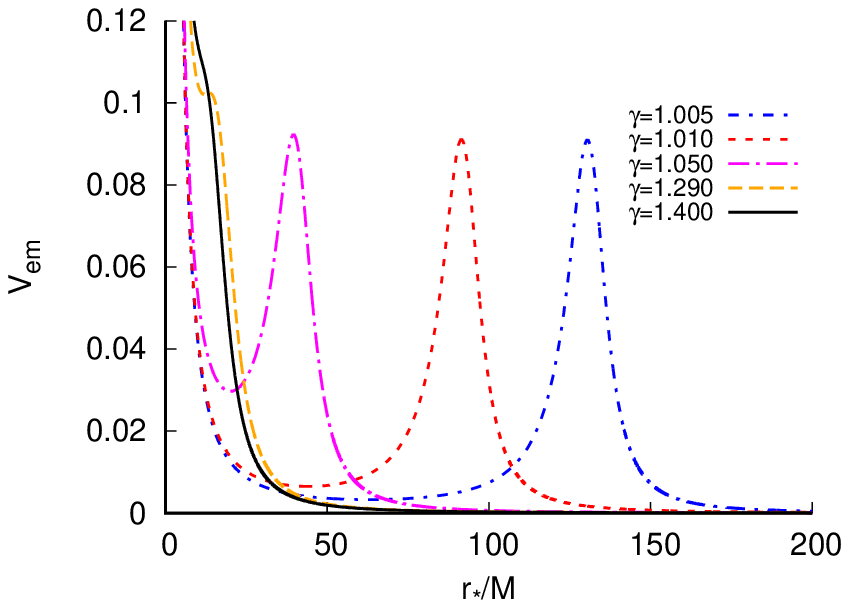}
\includegraphics[width=0.33\textwidth]{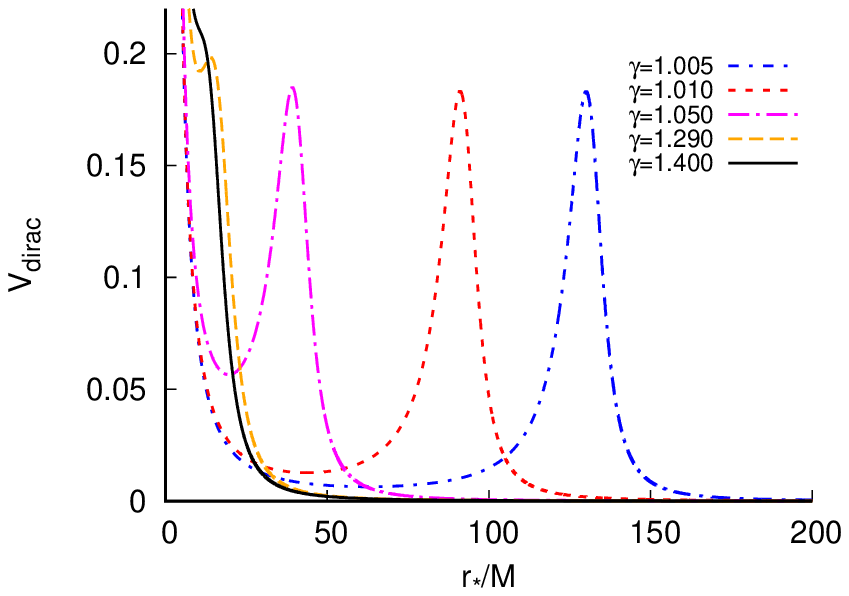}
\caption{Plots of the effective potential for massless scalar (left panel), electromagnetic (middle panel) and Dirac (right panel) perturbations with respect to the coordinate $r_*$ for $l=1$ and different values of $\gamma$ in the naked singularity regime.}
\label{fig:potential}
\end{figure*}

Figure~\ref{fig:potential} shows the effective potential for scalar, electromagnetic and Dirac perturbation as a function of the coordinate $r_*$ for different values of the dimensionless parameter $\gamma$ in the naked singularity regime.
\begin{figure*}[]
\includegraphics[width=0.45\linewidth]{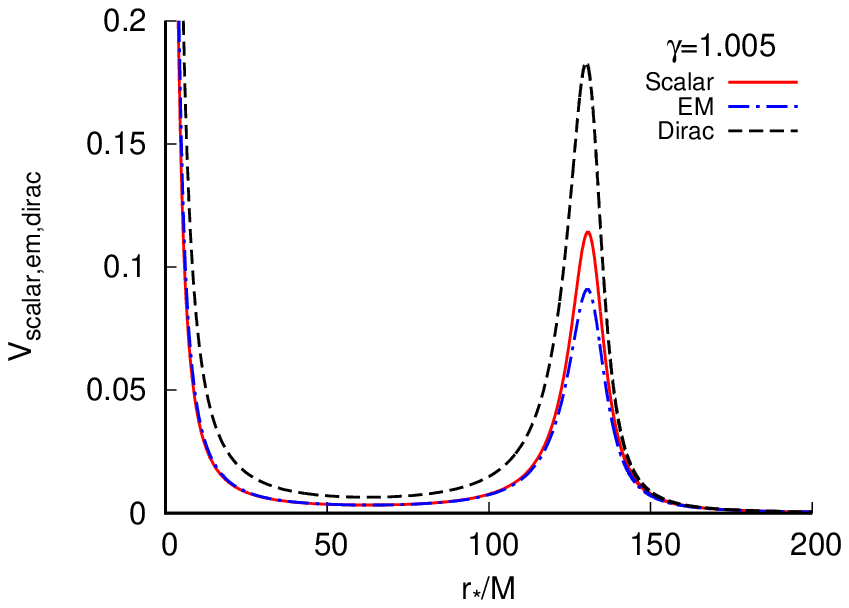}
\includegraphics[width=0.45\linewidth]{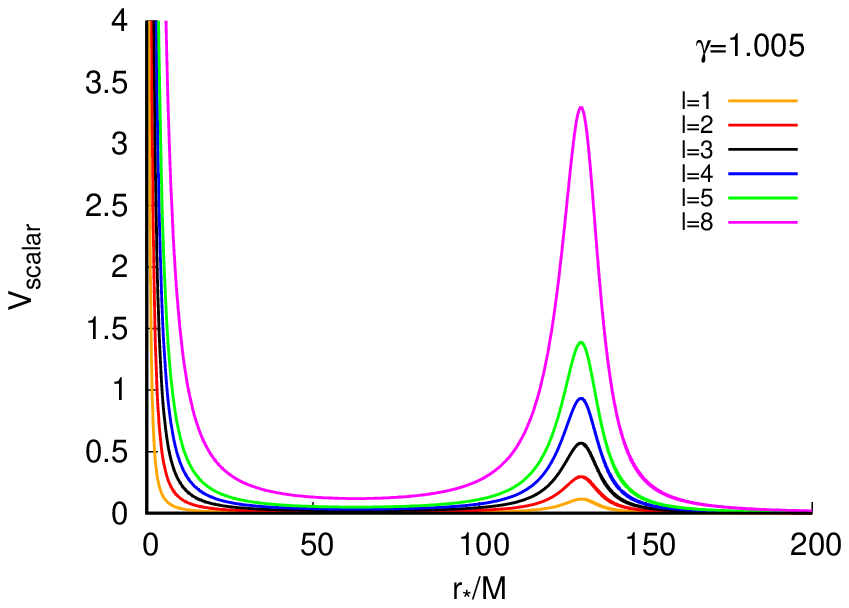}
\caption{The left panel shows the difference in the effective potential for the scalar, electromagnetic and Dirac perturbations for $\gamma=1.005$ and $l=1$. The right panel shows the effective potential for massless scalar perturbation for $\gamma=1.005$ for different values of $l$}
\label{fig:potential2}
\end{figure*}
We observe that in the regime of weakly naked singularity $\left( 1<\gamma<3\sqrt{3}/4\right)$, the potential profile for all the three types of perturbations are characterised by a peak at $r_*>0$ which rises to an infinite wall close to the singularity [$V(r\rightarrow 0)\rightarrow \infty$], except for the $l=0$ mode of scalar perturbation. The effective potential for the $l=0$ mode of scalar perturbation diverges to $-\infty$ close to the singularity, $\left[V_{scalar}^{l=0}\left(r \rightarrow 0\right) \longrightarrow - \infty\right]$, rendering the system unstable~\cite{konoplya2011RMP}. 
Henceforth, unless otherwise mentioned we will only consider $l>0$ modes of scalar and electromagnetic perturbations and $l\geq0$ modes of Dirac perturbation.

 The divergence of the effective potential distinguishes the spacetime~\eqref{eq:metric} with a naked singularity from the corresponding black hole solution in which case the effective potential is characterised by a single potential peak outside the event horizon. As $\gamma$ is increased from $\gamma\approx1$, the potential peak shifts towards smaller values of $r_*$,  until changes to a plateau and finally merges with the potential wall at sufficiently large values of $\gamma$. In this case, the effective potential is characterised solely by the infinite wall near the singularity. We also note from Fig.~\ref{fig:potential2} that the height of the peak of the potential profile for a given value of the parameter $\gamma$ changes with the type of perturbation considered. From Fig.~\ref{fig:potential2} we note that for a given $\gamma$ the height of the peak of the effective potential is maximum for the Fermionic (or Dirac) perturbation and minimum for electromagnetic perturbation. Also, for each type of perturbation the peak height and width increases with the multipole number. 
\section{Time evolution of Perturbation}\label{sec:timedomain}
To study the time-evoution of the perturbation we rewrite the perturbation equations~\eqref{eq:scalar_trts},\eqref{eq:em_trts} and \eqref{eq:dirac_trts} in terms of null coordinates, $u=t-r_*$ and $v=t+r_*$,
\begin{equation}\label{eq:td_uv}
4\frac{\partial^2}{\partial u \partial v}\psi_i\left(u,v\right) + V_{i}\left(u,v\right) \psi_i \left(u,v\right)=0~;~i \in (scalar,~ em~, dirac)
\end{equation}
To numerically integrate Eq.~\eqref{eq:td_uv}, we follow the integration scheme, proposed by Chirenti and Rezolla~\cite{chirenti2007CQG},
\begin{equation}\label{eq:discrt_scheme}
\psi_i(N)=\left(\psi_i(W)+\psi_i(E)\right)\frac{16 - \Delta^2 V_i(s)}{16 + \Delta^2 V_i(s)}-\psi_i(S)+\mathcal{O}\left(\Delta^4\right)
\end{equation}
where $\Delta$ is the step-size and $S=(u,v)$, $W=(u+\Delta,v)$, $E=(u,v+\Delta)$ and  $N=(u+\Delta,v+\Delta)$ are the grid-points in the $u-v$ plane. For an effective potential of the form, depicted Figs.~\ref{fig:potential},\ref{fig:potential2}, the above integration scheme is found to be more stable compared to more popular integration scheme due to Gundlach, Price and Pullin~\cite{gundlach1994PRD}, consistent with the observation in Ref.~\cite{chirenti2007CQG}.
In general, in the linear regime the eigenfrequencies are not sensitive to the choice of the initial conditions, hence, we model the initial perturbation by a Gaussian wave packet of width $\sigma$ centered around $v=v_c$~,
\begin{equation}
\psi_i (u=0,v)=e^{-\frac{\left( v-v_c \right)^2}{2 \sigma^2}}.
\end{equation}
We also assume that close to the singularity the perturbation is constant,
\begin{equation}\label{eq:cond_sng}
\psi_i (r_*=0,t)=\psi_i(u=v-v_0,v)= 0, ~ \forall t;~\epsilon<< 1.
\end{equation}
The choice of the null boundary condition~\eqref{eq:cond_sng} deserve some attention.

The presence of the central naked singularity renders the spacetime~\eqref{eq:metric} (with $\gamma>1$) globally non-hyperbolic. However, following Wald's suggestion~\cite{wald1980JMP} (see also Refs.\cite{horowitz1995PRD, ishibashi1999PRD, ishibashi2003CQG, helliwell2003GERG, ishibashi2004CQG, gibbons2004PTP, cardoso2006PRD}), it is possible to uniquely define dynamics of test fields even in such a spacetime, provided there exist a unique self adjoint extension of the operator $A_i$~,
\begin{equation}\label{eq:operatorA}
A_i\equiv-\frac{d^2}{dr_*^2} + V_i
\end{equation}
The operator $A_i$ acts on the Hilbert space of square integrable functions, $\mathcal{H}=L^2\left(r_*,dr_*\right)$ on a static hypersurface orthogonal to a unit time-like Killing vector $\partial_t$.  To analyse, the existence of an unique self adjoint extension of the operator $A_i$, one studies the eigenfunction of the equation,
\begin{equation}\label{eq:slfadjnt}
A_i \psi_i =\pm i~\psi_i~.
\end{equation}
The operator $A_i$ is said to be essentially self-adjoint (existence of a unique self adjoint extension) if atleast one of the eigenfunctions of $A_i$ (for each sign of $i$) fails to be square integrable near the singularity.
Close to the singularity, one can approximate 
\begin{eqnarray}
f(r) &\approx & 1-\sqrt{\frac{2}{\gamma  M}}~r^{1/2}+O\left(r^{3/2}\right),\\
r_* &\approx & r+ O\left(r^{3/2}\right) ,\\
V_i(r) &\approx & \frac{l (l+1) + 2C_i}{2 \gamma M^2 }+\frac{l (l+1)}{r^2}+O\left(r^{-3/2}\right),
\end{eqnarray}
where $C_i=1,~0,~3 (8 \gamma -1) (l+1)/(32 \gamma)$ for scalar, electromagnetic and Dirac perturbations respectively. Thus, close to the singularity one can write Eq.~\eqref{eq:slfadjnt} as 
\begin{equation}\label{eq:A_i}
 -\frac{d^2 \psi_i\left(r_*\right)}{dr_*^2} +\left(\frac{l (l+1)}{r_*^2}+ \cdots\right)\psi_i(r_*)=\pm i~ \psi_i\left( r_* \right)~.
\end{equation}
The general solution to Eq.~\eqref{eq:A_i} close to the singularity is of the form (for both signs of the eigenvalue $\pm i$)
\begin{equation}\label{eq:norm}
\psi_i\sim \mathcal{C}_1 \left( r_*^{-l} + \cdots \right) + \mathcal{C}_2 \left( r_*^{l+1} + \cdots \right)~\mbox{as}~ r_* \rightarrow 0.
\end{equation}
The first solution fails to be square integrable near the singularity and hence, $A_i$ is essentially self adjoint. It is important to note that addition of positive terms to the effective potential in Eq.~\eqref{eq:A_i} (including mass of the test field) does not alter the essential self adjointness of the operator $A_i$. Such terms effectively act as repulsive terms, increasing the rate of divergence of the larger solution and the convergence of the smaller solution close to the singularity~\cite{horowitz1995PRD, svitek2020AP}.
Further, assuming the time-dependence of the perturbation field as $\psi_i\left(t,r_*\right)=e^{-i \omega t} \psi_i\left(r_*\right)$, we write Eqs.~\eqref{eq:scalar_trts},\eqref{eq:em_trts} and \eqref{eq:dirac_trts} , near the singularity (upto leading order in $r_*$) as
\begin{equation}\label{eq:normsng}
-\frac{d^2 \psi_i\left(r_*\right)}{dr_*^2} +\left(\frac{l (l+1)}{r_*^2}+ \cdots\right)\psi_i(r_*)=\omega^2 \psi_i\left( r_* \right)~.
\end{equation}
The general solution of the Eq.~\eqref{eq:normsng} is of the form as Eq.~\eqref{eq:norm} and for $\psi_i$ to be normalizable close to the singularity, $\mathcal{C}_1$ must vanish, which implies
\begin{equation}
r_*^{l} \psi_i \mid_{r_*=0}~=0.
\end{equation} 
Thus, boundary condition~\eqref{eq:cond_sng} guarantees that the perturbation field is normalizable close to the singularity and is also consistent with Ref.~\cite{wald1980JMP}. 
\begin{figure*}[!]
\includegraphics[width=0.33\linewidth]{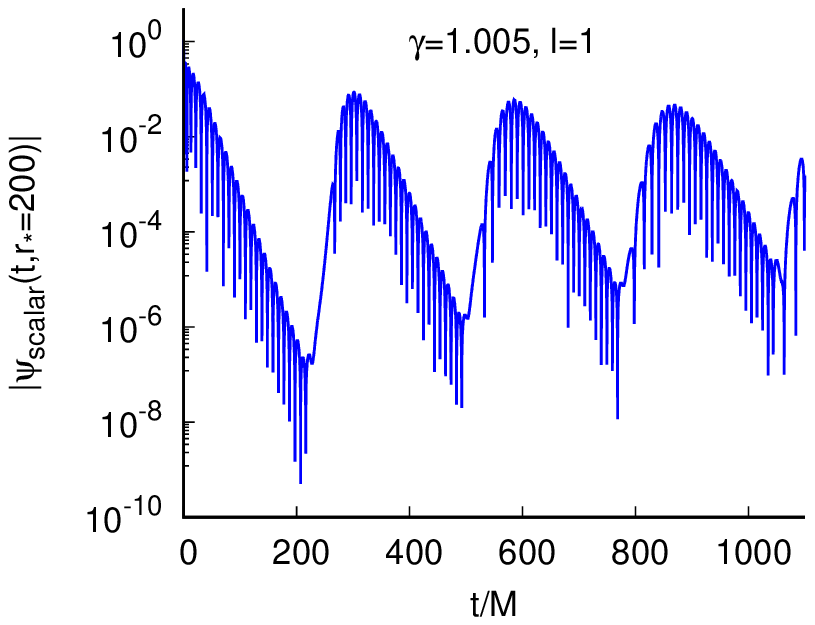}
\includegraphics[width=0.33\linewidth]{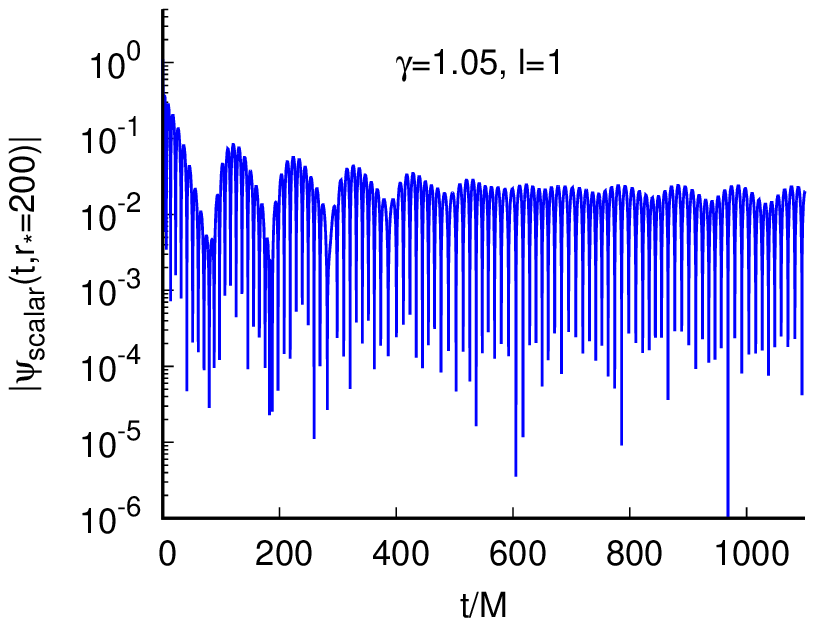}
\includegraphics[width=0.33\linewidth]{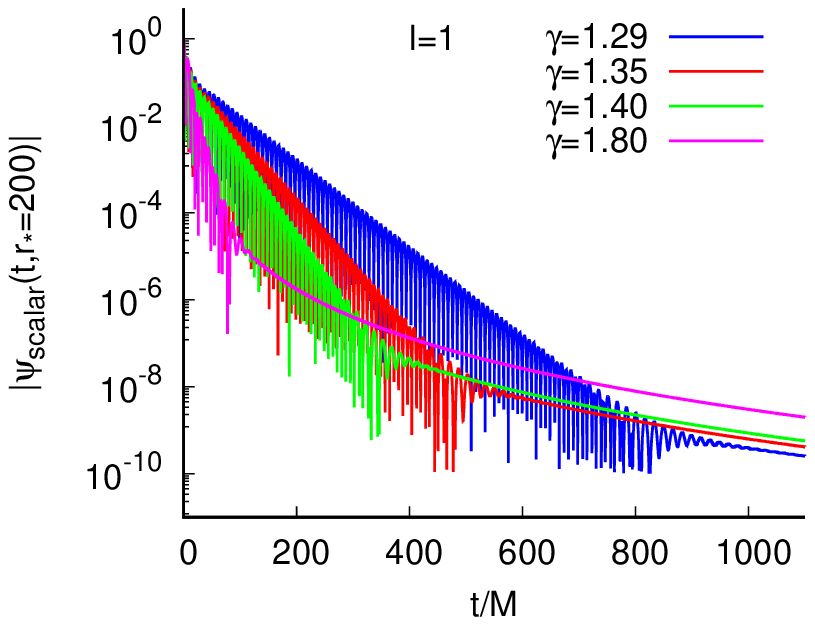}\caption{Semi-logarithmic plots of the time-evolution of massless scalar field perturbation
for the $l=1$ mode and different values of $\gamma$ . The time-profile has been extracted at $r_*=200$.}
\label{fig:TD_scalar}
\end{figure*}

\begin{figure*}[]
\includegraphics[width=0.33\linewidth]{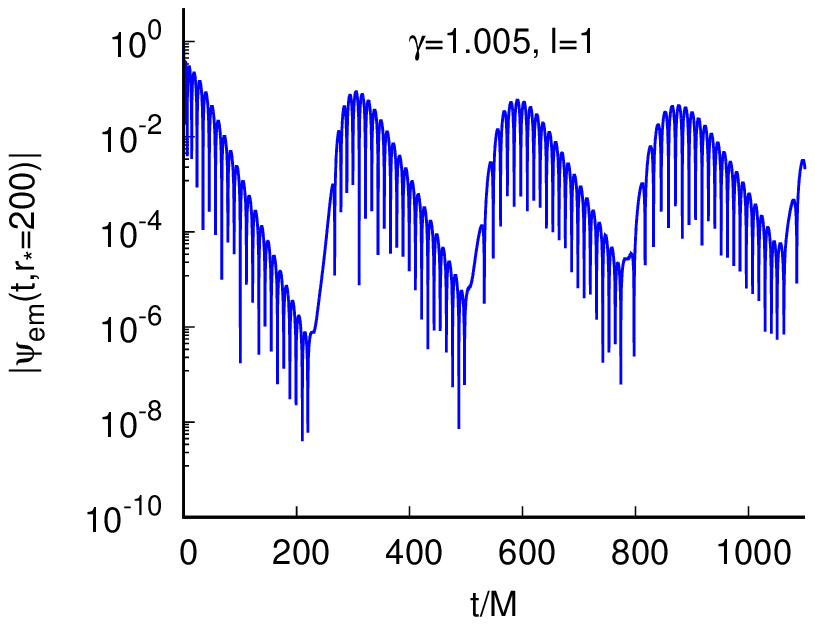}
\includegraphics[width=0.33\linewidth]{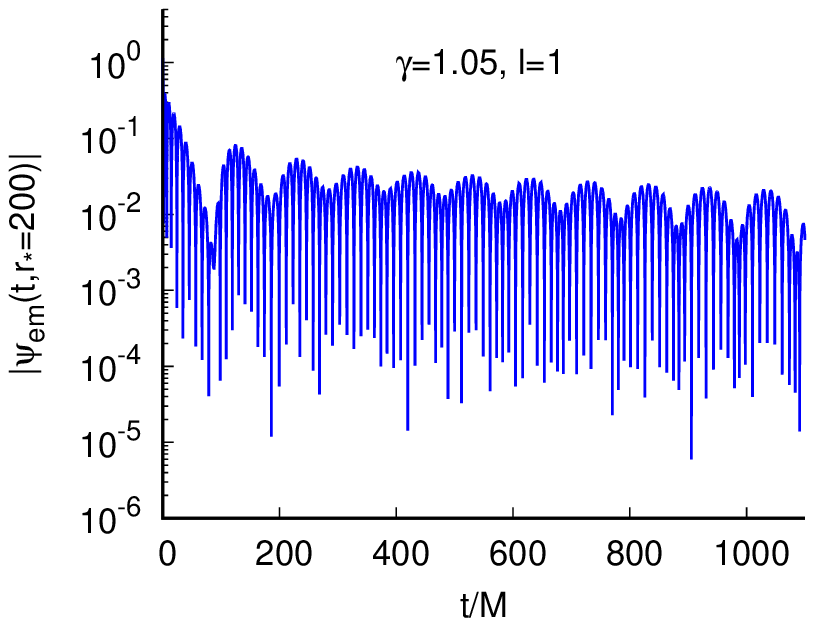}
\includegraphics[width=0.33\linewidth]{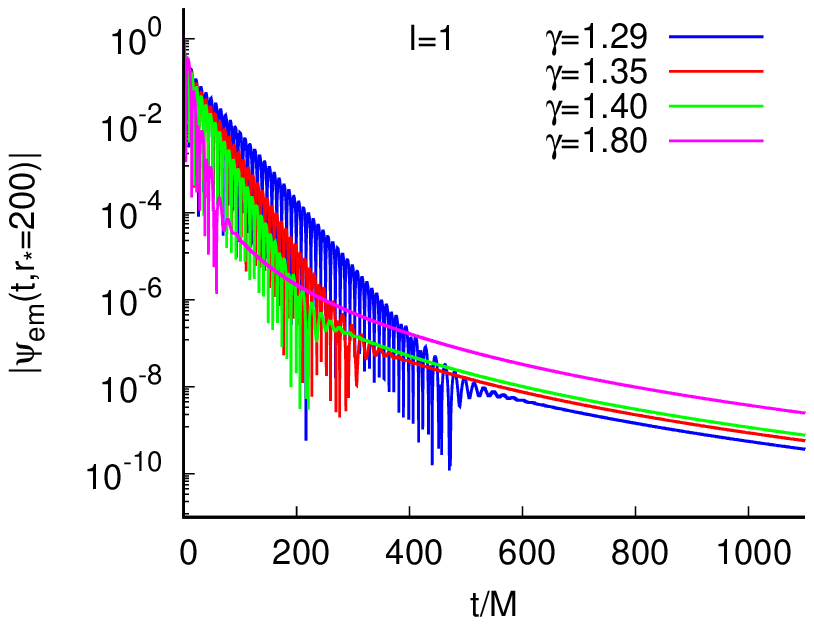}\\
\includegraphics[width=0.33\linewidth]{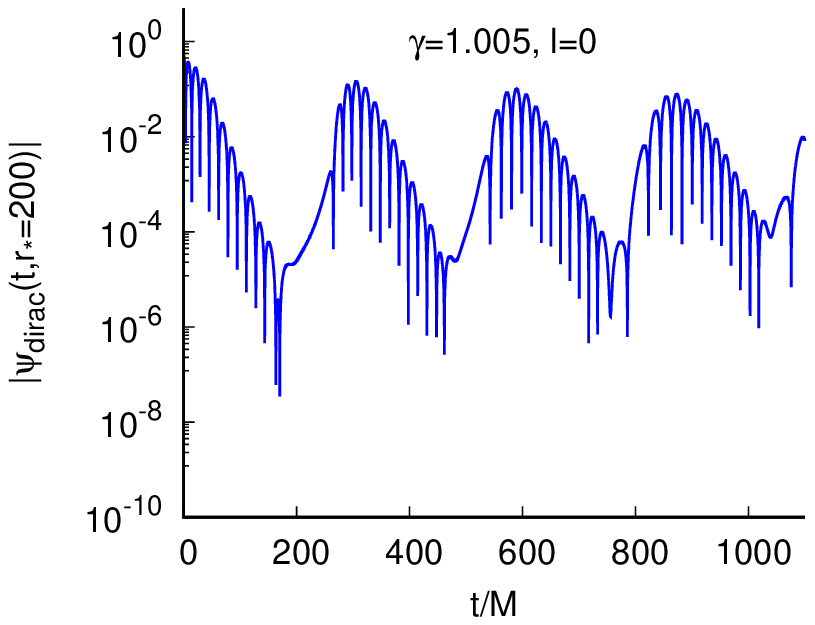}
\includegraphics[width=0.33\linewidth]{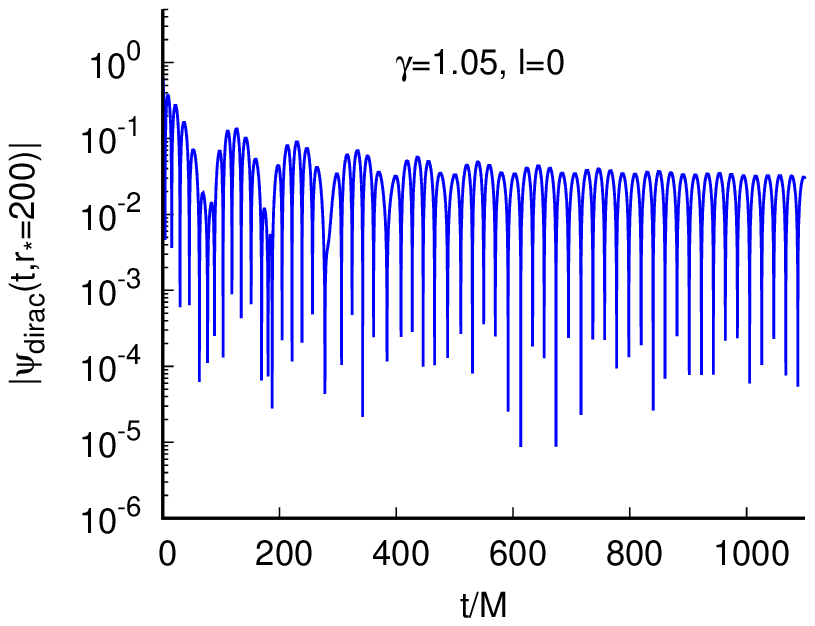}
\includegraphics[width=0.33\linewidth]{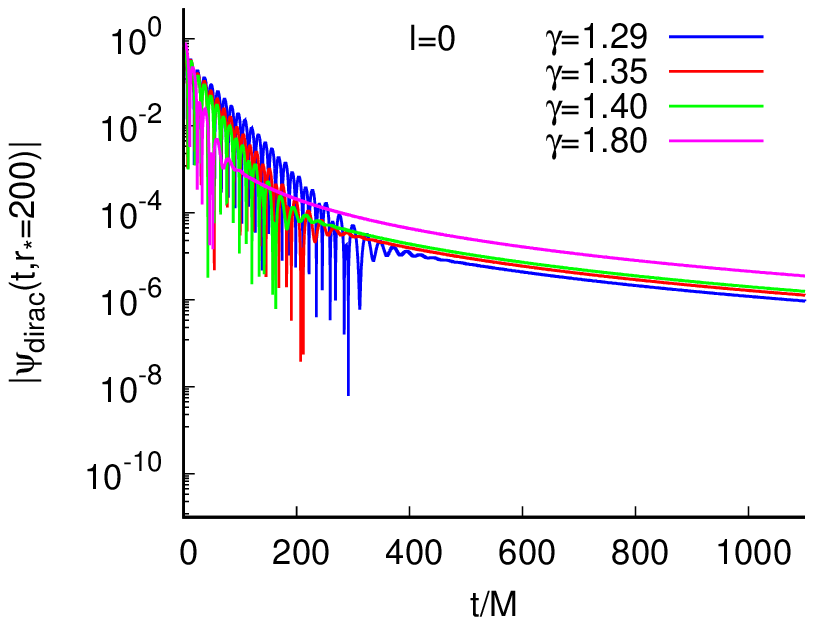}\\
\includegraphics[width=0.33\linewidth]{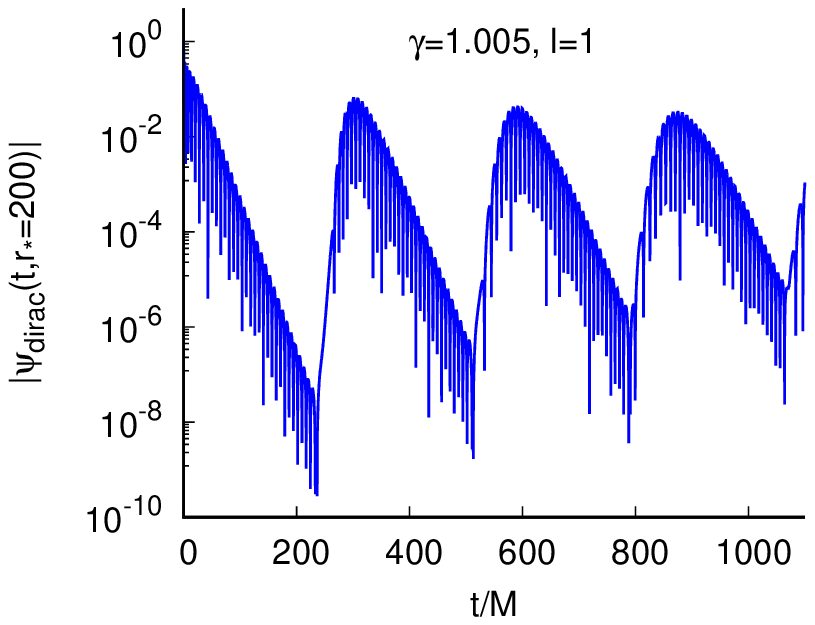}
\includegraphics[width=0.33\linewidth]{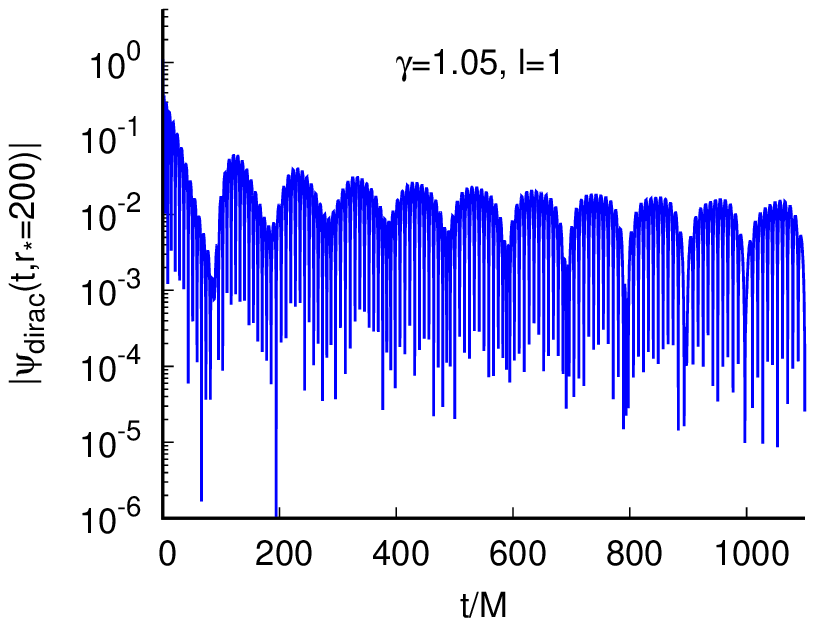}
\includegraphics[width=0.33\linewidth]{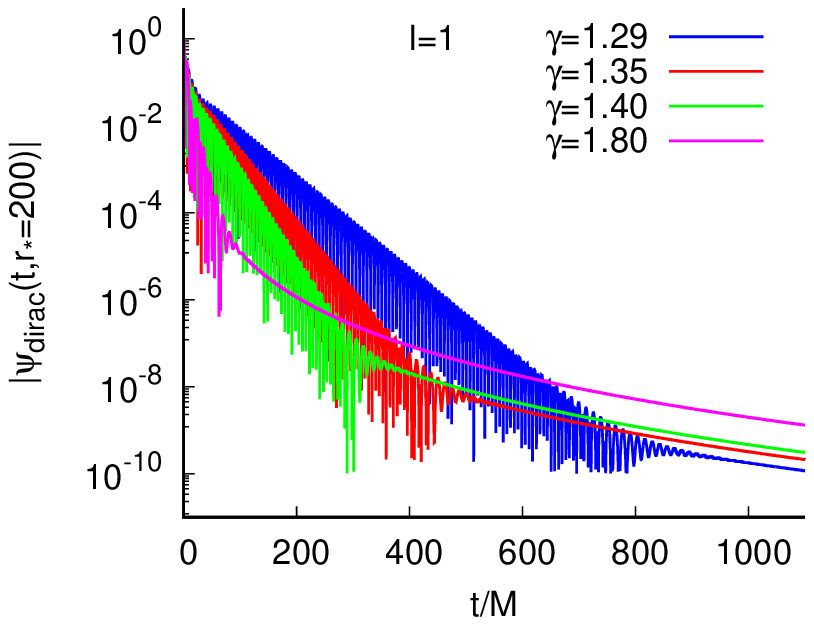}
\caption{Semilogarithmic plots of the time-evolution of the $l=1$ mode of electromagnetic (top panel) and  $l=0, 1$ modes of Dirac  perturbations (middle and bottom panel).
The time-profile has been extracted at $r_*=200$.}
\label{fig:TD_em-dirac_l1}
\end{figure*}

\begin{figure*}[]
\includegraphics[width=0.33\linewidth]{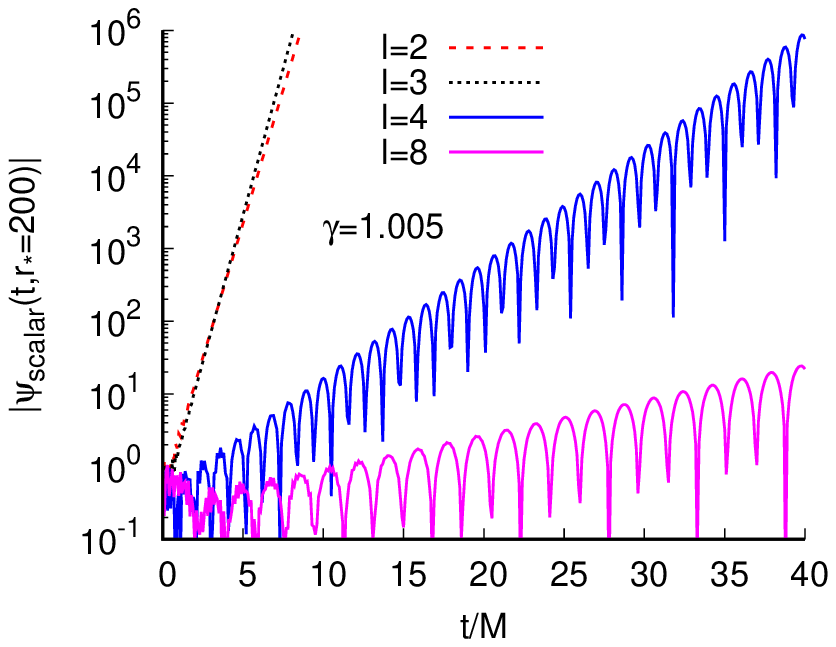}
\includegraphics[width=0.33\linewidth]{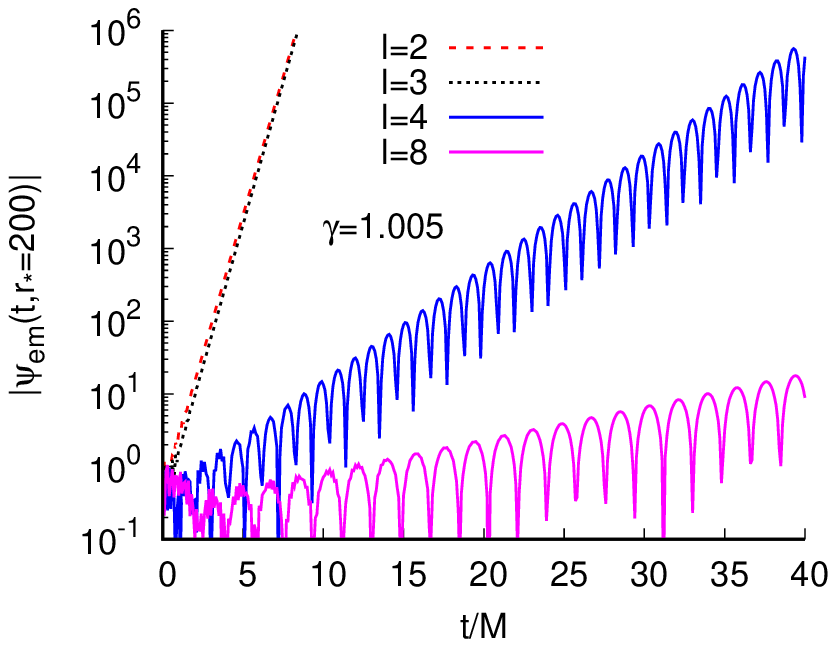}
\includegraphics[width=0.33\linewidth]{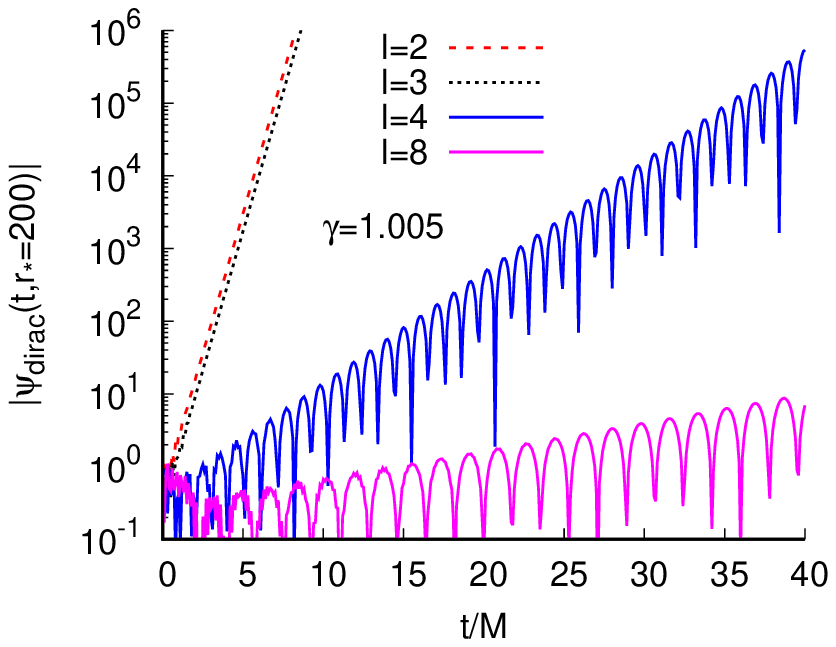}
\caption{Semilogarithmic plots of the time-evolution of the $l=2,3,4,8$ modes of scalar (left panel), electromagnetic (middle panel) and Dirac fields (right panel) for $\gamma=1.005$~.}
\label{fig:em-dirac_l2,4}
\end{figure*}

Figure~\eqref{fig:TD_scalar} shows time evolution of the scalar perturbation for the $l=1$ mode along a line of constant $r_*$. We observe that close to the black-hole limit ($\gamma=1$) the time profile of the scalar perturbation is characterised by distinct echoes with diminishing amplitude and frequency. As $\gamma$ increases, the echoes become  less prominent. For sufficiently large values of $\gamma$, the enveloping oscillation of the echoes align to give rise to characteristic quasinormal modes which later decays to give late-time tails. A study on the time domain profiles for all the three different types of perturbations suggests that the tail falls off as $~t^{-(2 l+ 2 +\gamma)}$.

 It may be noted that although the existence of distinct echoes in the time evolution of massless scalar perturbations are indicative of the presence of weakly naked singularity in the spacetime, the quasinormal ringing and late-time tails can also be observed in the time evolution of  massless scalar field, when the singularity is strongly naked.

Time evolution of the $l=1$ mode of electromagnetic and $l=0,1$ modes of Dirac fields also show similar characteristics (Fig.~\ref{fig:TD_em-dirac_l1}). 
However, for higher multipole modes of all three types of perturbation (scalar, electromagentic and Dirac) grows unboundedly with time suggesting an instability. Figure~\ref{fig:em-dirac_l2,4} shows the time-evolution of the $l=2, 3, 4, 8$ modes of the perturbations. To ascertain that the instability of the $4D$-EGB naked singularity-spacetime is not a numerical artefact, we have also checked our results by slightly shifting the location of the inner boundary, $\left(r_{in}=r_{in}^0\pm \epsilon^2,~\mbox{where}~r_{in}^0\sim \epsilon;~  \epsilon<<1 ~\right)$. 

\begin{figure*}[!]
\includegraphics[width=0.495\linewidth]{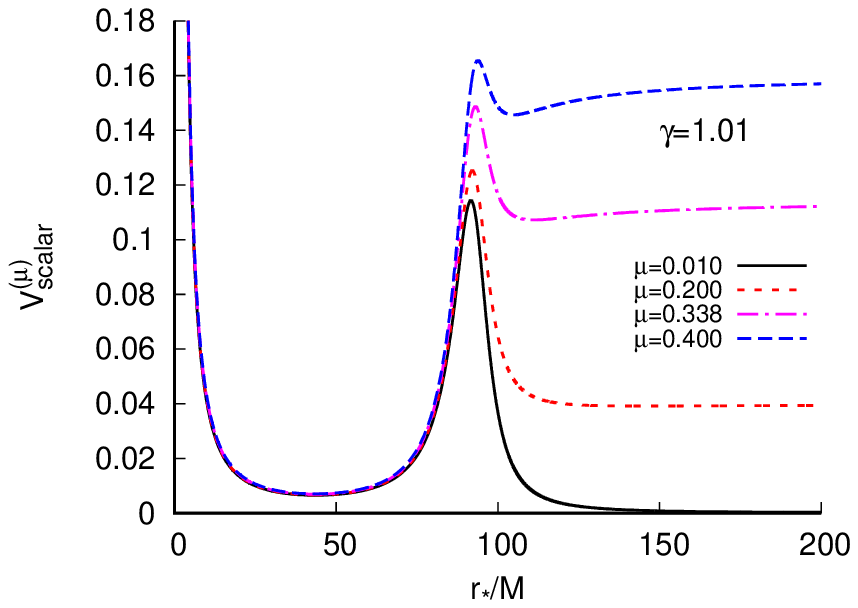}
\includegraphics[width=0.495\linewidth]{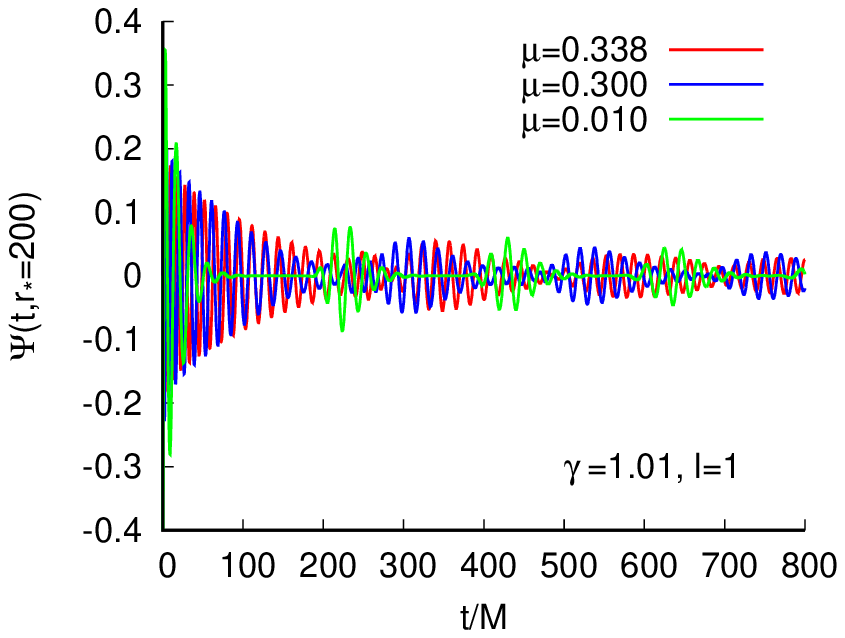}
\caption{Plots of the effective potential for massive scalar field (left) and the corresponding time-domain profile for the $l=1$ mode with $\gamma=1.005$.}
\label{fig:massive}
\end{figure*}

Once the echoes align, mode frequencies can be extracted from the time profile by using Prony's method of fitting the time-domain data with a series of damped exponentials with some excitation factors~\cite{konoplya2011RMP, berti2007PRD},
\begin{equation}\label{eq:prony_ser}
\psi\left(t\right) \simeq \sum_{j=1}^{p} C_j e^{-i \omega_j t}~,
\end{equation}
where $\omega_j$ is the complex quasinormal frequency of the $j$th mode. The real part of the quasinormal normal frequency corresponds to the actual frequency of the wave motion while the imaginary part corresponds to the damping rate. The fundamental quasinormal mode frequency is characterised by the value of $\omega_j$ with the lowest damping rate i.e., with the smallest $Im(\omega)$. Table~\ref{tab:qnm} shows the characteristic fundamental quasinormal frequencies for the $l=1$ mode of scalar and electromagnetic perturbation and $l=0,1$ modes of Dirac perturbation. The quasinormal frequencies has been extracted for values of the dimensionless parameter $\gamma$ for which the echoes have aligned. We observe that the magnitude of both the real and imaginary parts of the   quasinormal frequencies  increases with $\gamma$ for each type of perturbation.  
If we consider the mass ($\mu$) of the perturbing scalar field to be non zero then the effective potential in Eq.~\eqref{eq:v_scalar} gets modified to,
\begin{equation}
V_{scalar}^{(\mu)}(r)=f(r)\left(  \frac{l(l+1)}{r^2} + \frac{1}{r} \frac{d f(r)}{d r}+ \mu^2 \right)~.
\end{equation}
Thus the asymptotic value of the effective potential changes to $V^{(\mu)}_{scalar}\left(r\to \infty\right) \rightarrow \mu^2$. For sufficiently large mass of the probing scalar field, there exists a trough in the effective potential outside the peak, resulting in quasi-bound states, which are manifested as elongation of the individual echoes (Fig.~\ref{fig:massive}).

The quasinormal mode frequencies evaluated using Prony's method depends on the choice of the starting point of the ringdown profile. To eliminate possible errors in the determination of the quasinormal mode frequencies we have verified the quasinormal mode frequencies with time-profile date generated using different grid sizes $\left(\Delta\right)$.  For each such time profile, we have checked the stability of the fundamental quasinormal mode frequencies by fitting with the series in Eq.~\eqref{eq:prony_ser} with different number of terms ($\sim 100 - 200$).
\begin{table*}[!]
\centering
\caption{Characteristic fundamental quasinormal frequencies for $l=1$ mode of massless scalar and electromagnetic perturbations and $l=0,1$ modes of massless Dirac perturbations. }
\label{tab:qnm}
\begin{tabular*}{\textwidth}{c @{\extracolsep{\fill}} cccc}
\hline\hline
\multirow{2}{*}{$\gamma$} &
  \multirow{2}{*}{\begin{tabular}[c]{@{}c@{}}Scalar \\ $\omega~(l=1)$\end{tabular}} &
  \multirow{2}{*}{\begin{tabular}[c]{@{}c@{}}Electromagnetic\\ $\omega~(l=1)$\end{tabular}} &
  \multicolumn{2}{c}{Dirac} \\ \cline{4-5} 
       &                     &                      & $\omega~(l=0)$      & $\omega~(l=1)$       \\ \hline
$1.25$ & $0.3699 - 0.0082 i$ & $0.3528 - 0.0132 i$ & $0.2545 - 0.0131 i$ & $0.4676 - 0.0084 i$ \\
$1.28$ & $0.3797 - 0.0107 i$ & $0.3613 - 0.0166 i$  & $0.2614 - 0.0149 i$ & $0.4787 - 0.0113 i$ \\
$1.29$ & $0.3827 - 0.0116 i$ & $0.3638 - 0.0177 i$  & $0.2634 - 0.0155 i$ & $0.4820 - 0.0123 i$ \\
$1.30$  & $0.3855 - 0.0125 i$ & $0.3662 - 0.0188 i$  & $0.2653 - 0.0161 i$ & $0.4852 - 0.0133 i$ \\
$1.35$ & $0.3977 - 0.0170 i$ & $0.3766 - 0.0243 i$  & $0.2750 - 0.0207 i$ & $0.4985 - 0.0186 i$ \\
$1.40$  & $0.4074 - 0.0215 i$ & $0.3842 - 0.0301 i$  & $0.2827 - 0.0248 i$ & $0.5097 - 0.0241 i$ \\
$1.60$  & $0.4314 - 0.0390 i$ & $0.3943 - 0.0451 i$  & $0.2957 - 0.0392 i$ & $0.5375 - 0.0399 i$ \\ \hline
\end{tabular*}%
\end{table*}


\section{Conclusion} \label{sec:conclusion}
The Cosmic Censorship Conjecture suggests that spacetime singularities must always be hidden by an event horizon, however, it has been argued that under suitable initial condition gravitational collapse may lead to a naked singularity~\cite{penrose1969RNC}.  In general for gravitational collapse, the quantum considerations are towards an avoidance of a singularity \cite{harada2001PRD, bojowald2005PRL, liu2014PRD, kiefer2019PRD}. So, if for certain values of theory parameters, a gravity theory predicts the occurrence of a naked singularity, then it is of paramount importance to check the stability of such a spacetime with naked singularity against perturbation. If such a spacetime happens to be sufficiently stable, then one asks the associated question of how to observationally distinguish such an atypical spacetime. 

In the present work, we considered an asymptotically flat, static, spherically symmetric spacetime~\eqref{eq:metric} with a central singularity. We observed that the singularity becomes globally naked for $\gamma>1$. It is important to emphasise that the metric~\eqref{eq:metric} satisfies the field equations of all variants of the (consistent) $4D$-EGB theory~\cite{lu2020PLB, kobayashi2020JCAP, fernandes2020PRD, hennigar2020JHEP, aoki2020PLB}, hence, we studied the stability and response of such a naked singularity-spacetime against perturbation by test fields without  resorting to any particular version of the (consistent) $4D$-EGB theory. 

We added test scalar, electromagnetic and Dirac fields in the background of the $4D$-EGB naked singularity-spacetime and observed the time evolution of the perturbations numerically. The effective potential of all the three types of perturbation diverges to $\infty$ close to the singularity. So, we chose the null Dirichlet boundary condition consistent with~\cite{chowdhury2020PRD,chirenti2013PRD}. We observed that for $l=1$ modes of scalar, electromagnetic perturbation and $l=0,1$ modes of Dirac perturbation the time-domain profile give rise to distinct echoes the dimensionless parameter $\gamma$ is slightly greater than unity (weakly naked singularity regime). As gamma is increased the timegap between the individual echoes decreases and finally for sufficiently large gamma the echoes align to yield characteristic quasinormal frequency of the spacetime. However, as $l$ is increased from unity, the time-domain profile~(Fig.~\ref{fig:em-dirac_l2,4}) suggests an instability. We have verified the instability of the spacetime till $l=10$. 

So, we conclude that the $4D$-EGB spacetime with naked singularity is unstable against test scalar, electromagnetic and Dirac perturbations which constraints the Gauss-Bonnet coupling constant $\alpha \leq M^2$.
We plan to extend our analysis to study the full gravitational perturbation of the $4D$-EGB black hole and naked singularity in all versions of the (consistent) $4D$-EGB theory. Such an analysis will not only provide an opportunity to constraint the parameter space of $\alpha$ but will also be able to predict direct observational distinctions between the  different versions of the $4D$-EGB theories.
\begin{acknowledgements}
AC acknowledges the use of the ``Dirac" computing facility of IISER Kolkata.
\end{acknowledgements}
\vspace{1cm}
\def\bibsection{}  
\centerline{\small {\textbf{REFERENCES}} }
\bigskip
\bibliography{biblio} 
\end{document}